\documentclass[aps,pre,preprint,12pt]{revtex4-2}

\usepackage[utf8]{inputenc}
\usepackage[T1]{fontenc}

\usepackage{amsmath}
\usepackage{amsfonts}
\usepackage{amssymb}
\usepackage{bm}
\usepackage{mathrsfs}

\usepackage{graphicx}
\usepackage{subcaption}
\usepackage{float}
\usepackage{indentfirst}

\usepackage{xcolor}
\usepackage[colorlinks=true,
            citecolor=blue,
            linkcolor=blue,
            urlcolor=blue,
            filecolor=blue]{hyperref}

\begin{document}

\title {Repulsive barrier suppression of ionization stabilization}

\author{Murilo D. Forlevesi}
\email{murilo.deliberali@unesp.br}
\affiliation{Departamento de Física, UNESP - Universidade Estadual Paulista, Rio Claro SP, 13506-900, Brazil}
\author{Gabriel Albertin Amici}
\author{Emanuel Fernandes de Lima}
\email{eflima@ufscar.br}
\affiliation{Departamento de F\'isica, Universidade Federal de S\~ao Carlos (UFSCar)\\ S\~ao Carlos, SP 13565-905, Brazil}

\date{\today}

\begin{abstract}

Ionization stabilization is a well-known phenomenon in strongly driven Soft-Coulomb atomic models, where the ionization probability decrease as the field amplitude increases. In this work, we investigate how this process is affected by introducing a repulsive Morse barrier into the binding potential, leading to the Morse-Soft-Coulomb (MsC) model. A systematic comparison between the Soft-Coulomb and Morse-Soft-Coulomb systems is performed for different values of the softening parameter. Ionization probabilities, escape-time and Lagrangian descriptor maps reveal that the stabilization observed in the Soft-Coulomb model is suppressed in the Morse-Soft-Coulomb system. To elucidate the origin of this behavior, we analyze the corresponding Kramers-Henneberger effective potentials. While the Soft-Coulomb model develops a symmetric double-well structure supporting a large trapping region, the Morse-Soft-Coulomb potential exhibits a single effective minimum with a significative smaller trapping region. Our result indicate that the repulsive barrier leads to the suppression of the ionization stabilization. 

\end{abstract}

\maketitle
\newpage
\section{Introduction}

One-dimensional model potentials have long served as valuable theoretical laboratories for investigating atomic and molecular dynamics \cite{Sethi2009,Forlevesi2018,DeLima2013,Loudon2016,Burke2013,Korolkov1996,Menis1992,r6hy-vkxh}. Their mathematical simplicity allows detailed classical and quantum analyses while preserving the essential mechanisms underlying phenomena such as photoassociation, photodissociation, strong-field ionization, and electron transport \cite{Protopapas1997,Carrillo2020,Burke2009,Jones1993}. In particular, these models provide a natural framework for understanding how the shape of the binding potential influences phase-space organization and the resulting dynamical behavior under external laser driving \cite{forlevesi2021nonlinear,de2020nonchaotic}.

The interaction of atoms and molecules with intense laser fields has revealed a variety of nonlinear phenomena that challenge simple intuition about ionization dynamics. Among them, ionization stabilization remains one of the most remarkable examples, corresponding to a counterintuitive regime in which the ionization probability decreases as the laser intensity increases~\cite{PhysRevLett.64.862,Pont1990,gavrila1984free,gavrila2005atomic,Potvliege1989,joachain1994theory,Grobe,Kulander,PhysRevLett.68.170}. Since its first theoretical prediction, stabilization has been extensively investigated in both quantum and classical frameworks and remains a paradigmatic phenomenon in strong-field physics~\cite{Faisal1987,Joachain2012,Becker2012,PhysRevA.91.023406}.

From a quantum-mechanical perspective, ionization stabilization has traditionally been interpreted in terms of laser-dressed states within the Kramers--Henneberger (KH) framework, in which the electron is described in an oscillating reference frame and experiences an effective time-averaged binding potential \cite{Henneberger1968,Gavrila2002}. Within this framework, the formation of effective trapping regions can substantially suppress ionization even in the presence of extremely strong fields \cite{davier1968rho,Bray}. From a classical perspective, stabilization has long been associated with phase-space structures that inhibit electron escape despite strong external driving \cite{Mauger2009,Chandre2010,cvitanovic2005chaos,Richter_2013,PhysRevA.46.1638,A_Lopez-Castillo_1999,PhysRevA.48.746}. More recently, these two viewpoints have been unified by demonstrating that KH states originate from periodic orbits and their associated phase-space structures in the full time-dependent Hamiltonian. This dynamical interpretation establishes a direct connection between the KH picture, classical invariant structures, and ionization stabilization, providing a unified description of electron trapping in intense laser fields \cite{Floriani2024KHScars,floriani2022bogolyubov,ivanov2022interference}.

One-dimensional Soft-Coulomb potentials have played a central role in the investigation of stabilization phenomena \cite{Su1990,Dziubak2010}. These model potentials retain the correct long-range Coulomb behavior while regularizing the singularity at the origin, thereby providing numerically stable and computationally efficient models for large-scale simulations of strong-field dynamics \cite{Javanainen1988,Grobe1993}. Numerous numerical and theoretical studies have demonstrated that one-dimensional Soft-Coulomb models exhibit pronounced stabilization windows over broad ranges of laser intensities and driving frequencies, making them a standard benchmark for investigating the dynamical mechanisms underlying ionization stabilization \cite{Reiss1992,Schwengelbeck1994,Baisong_1998,norman2015nonlinear}. However, previous studies have largely associated the absence of stabilization with the singular character of the Coulomb potential \cite{T_Menis_1992}. Our results suggest a different interpretation, indicating that the decisive factor is the presence of a repulsive barrier, which profoundly modifies the phase-space structures responsible for electron trapping. In this view, the Coulomb singularity may act through the repulsive dynamics it induces, rather than being the fundamental mechanism responsible for suppressing stabilization \cite{Carrillo-Bernal_2020}.

In this work, we investigate the recently proposed Morse-Soft-Coulomb (MsC) model, obtained by replacing the negative-coordinate branch of the Soft-Coulomb potential by a Morse repulsive barrier \cite{Lima2025}. This modification generates a strongly asymmetric binding potential while preserving the Soft-Coulomb interaction on the positive side. We compare the ionization dynamics of the Soft-Coulomb and Morse-Soft-Coulomb systems under identical driving conditions and investigate how this modification affects the stabilization process.

Our results reveal a striking qualitative difference between the two models. While the Soft-Coulomb potential displays the well-known stabilization regime, the Morse-Soft-Coulomb system exhibits a strong suppression of this behavior. To understand the origin of this effect, we analyze ionization probabilities, escape-time and Lagrangian descriptors maps and effective Kramers-Henneberger potentials. The combined analysis shows that the disappearance of stabilization is accompanied by a profound reorganization of the transport mechanisms governing ionization. In particular, the phase-space structures associated with long trapping times in the Soft-Coulomb model are strongly modified by the introduction of the Morse repulsive branch.

The paper is organized as follows. In Sec.~II we introduce the Soft-Coulomb and Morse-Soft-Coulomb models. In Sec.~III we describe the numerical procedures employed in the simulations, present the ionization probabilities and discuss the suppression of stabilization in the Morse-Soft-Coulomb system. Escape-time and Lagrangian descriptors maps and Kramers-Henneberger effective potentials are analyzed in Sec.~IV. Finally, the main conclusions are summarized in Sec.~V.

% ============================================================
\section{Model}
% ============================================================

We consider a one-dimensional classical electron driven by an external laser field. The dynamics is governed by the Hamiltonian

\begin{equation}
H(r,p,t)=\frac{p^2}{2}+V(r)+rF(t),
\label{eq:hamiltonian}
\end{equation}
where $r$ and $p$ are the electron coordinate and momentum, respectively, and $F(t)$ is the time-dependent electric field.

Two model potentials are compared. The first one is the standard Soft-Coulomb (SC) potential,

\begin{equation}
V_{\rm SC}(r)
=
-\frac{1}{\sqrt{r^2+\alpha^2}},
\label{eq:soft_coulomb}
\end{equation}
where $\alpha$ is the softening parameter.

The second one is a Morse-Soft-Coulomb (MsC) potential constructed by combining a Soft-Coulomb branch for positive coordinates with a Morse repulsive branch for negative coordinates,

\begin{equation}
V_{\rm MsC}(r)
=
\begin{cases}
V_{\rm SC}(r),
&
r>0,
\\[10pt]
D_e\left[1-\exp\left[-\beta(r-r_e)\right]\right]^2-D_e,
&
r\le 0.
\end{cases}
\label{eq:msc}
\end{equation}

Here, $D_e$ is the Morse well depth, $\beta$ controls the spatial range of the Morse interaction, and $r_e$ is the equilibrium position. In the calculations, we set $D_e=1/\alpha$, $\beta=1/(\alpha\sqrt{2})$, $r_e=0$. With this choice, both branches have the same value at the origin, $V_{\rm SC}(0)=V_{\rm MsC}(0)=-1/\alpha$, while the MsC potential is continuous up to the second derivative at the origin. The comparison between SC and MsC therefore isolates the effect of replacing one side of the Soft-Coulomb potential by a Morse-type repulsive barrier. Throughout this work we consider two values of the softening parameter, $\alpha=1$ and $\alpha=\sqrt{2}$ since these are the most used values found in the literature \cite{norman2015nonlinear,MProtopapas_1997}.

The external field is written as

\begin{equation}
F(t)=F_0 f(t)\sin(\omega t),
\label{eq:field}
\end{equation}
where $F_0$ is the field amplitude, $\omega$ is the carrier frequency, and $f(t)$ is the pulse envelope. The envelope consists of a linear turn-on, a constant plateau, and a linear turn-off,

\begin{equation}
f(t)=
\begin{cases}
\dfrac{\omega t}{12\pi},
&
0\le t\le t_{\rm on},
\\[8pt]
1,
&
t_{\rm on}<t\le t_{\rm on}+t_{\rm flat},
\\[8pt]
\dfrac{102\pi-\omega t}{2\pi},
&
t_{\rm on}+t_{\rm flat}<t\le t_{\rm total},
\\[8pt]
0,
&
t>t_{\rm total}.
\end{cases}
\label{eq:envelope}
\end{equation}

The pulse parameters are $t_{\rm on}=6T,t_{\rm flat}=44T,t_{\rm total}=51T$, where $T$ is the oscilation period. The pulse shape employed in this work follows the standard protocol widely adopted in previous investigations of ionization stabilization in Soft-Coulomb systems. In particular, the choice of a carrier frequency $\omega=0.8$, together with a smooth turn-on, an extended plateau region, and a gradual turn-off, closely follows the pulse configurations used in several classical and quantum studies of stabilization. Such a choice facilitates direct comparison with earlier results reported in the literature and ensures that any observed differences can be attributed to modifications of the binding potential rather than to changes in the laser pulse shape \cite{Grobe,T_Menis_1992,norman2015nonlinear}.

% ============================================================
\section{Results and Discussions}
% ============================================================

\subsection{Ionization stabilization in the Soft-Coulomb and Morse-Soft-Coulomb models}

Ionization probabilities are computed from ensembles of trajectories initialized on the field-free energy shell $E_i=-0.5$. Two complementary criteria are considered to classify ionizing trajectories. According to the distance criterion, a trajectory is ionized if its final position satisfies $|r(t_f)|>20$ a.u., whereas according to the energy criterion, ionization occurs when the final mechanical energy is positive, $E_{\mathrm{mech}}(t_f)>0$. Previous studies have shown that both criteria produce qualitatively similar ionization probabilities and lead to the same physical conclusions regarding the underlying dynamics \cite{norman2015nonlinear}. Unless otherwise stated, the results discussed in this work are based on the asymptotic energy criterion. The ionization probability is then defined as

\begin{equation}
P_{\rm ion}=\frac{N_{\rm ion}}{N_{\rm tot}},
\end{equation}
where $N_{\rm ion}$ is the number of ionized trajectory.

\begin{figure}[!t]
\centering

\begin{subfigure}{0.7\textwidth}
\includegraphics[width=\linewidth]{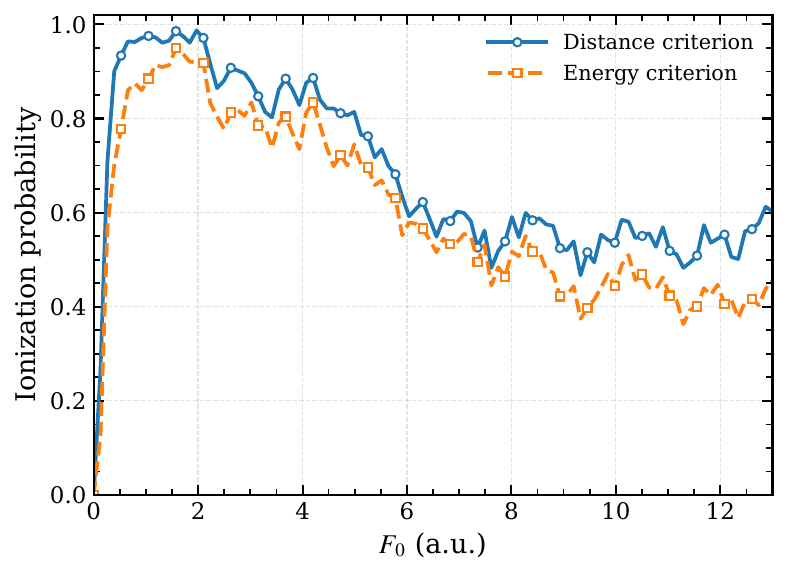}
\caption{Soft-Coulomb}
\end{subfigure}
\hfill
\begin{subfigure}{0.7\textwidth}
\includegraphics[width=\linewidth]{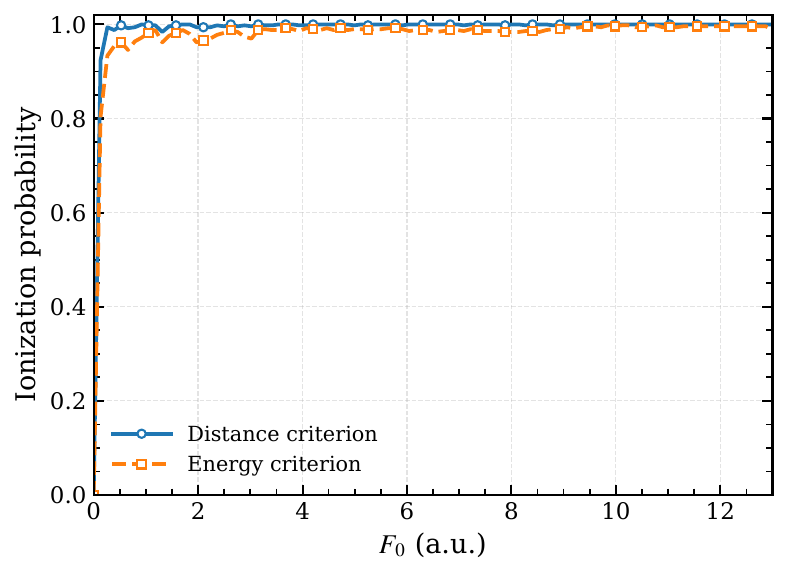}
\caption{Morse-Soft-Coulomb}
\end{subfigure}
\caption{
Ionization probability as a function of the field amplitude $F_0$ for
(a) the Soft-Coulomb and (b) the Morse--Soft-Coulomb (MsC) potentials,
with $\alpha=1$ and initial energy $E_i=-0.5$. The curves correspond to
the distance- and energy-based ionization criteria.
}
\label{fig:probability_alpha1}
\end{figure}

\begin{figure}[!t]
\centering
\begin{subfigure}{0.7\textwidth}
\includegraphics[width=\linewidth]{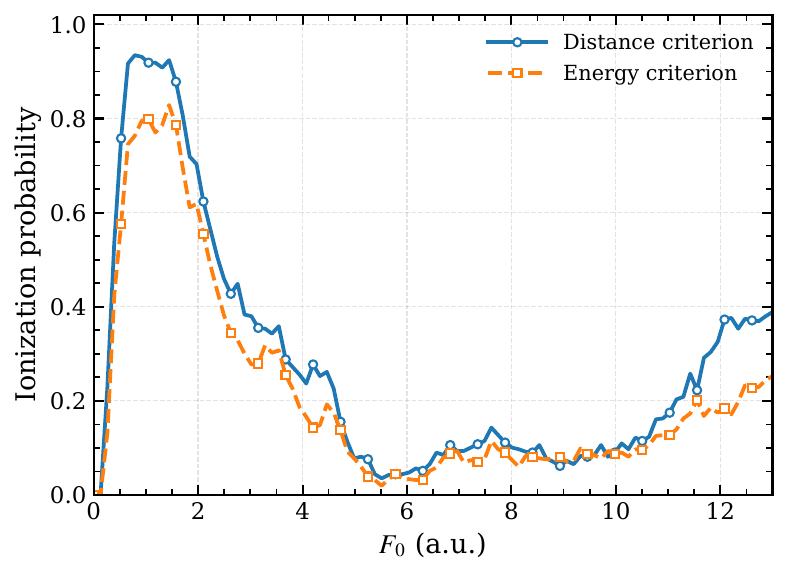}
\caption{Soft-Coulomb}
\end{subfigure}
\hfill
\begin{subfigure}{0.7\textwidth}
\includegraphics[width=\linewidth]{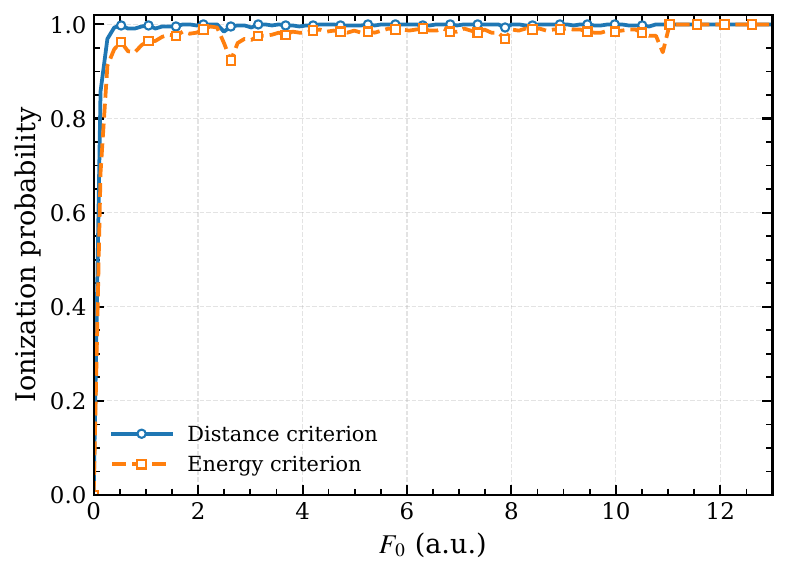}
\caption{Morse-Soft-Coulomb}
\end{subfigure}
\caption{
Ionization probability as a function of the field amplitude $F_0$ for
(a) the Soft-Coulomb and (b) the Morse--Soft-Coulomb (MsC) potentials,
with $\alpha=\sqrt 2$ and initial energy $E_i=-0.5$. The curves correspond to
the distance- and energy-based ionization criteria.}
\label{fig:probability_alpha_sqrt2}
\end{figure}

Figures~\ref{fig:probability_alpha1} and \ref{fig:probability_alpha_sqrt2} compare the ionization probabilities obtained for the Soft-Coulomb and Morse-Soft-Coulomb models as a function of the field amplitude for $\alpha=1$ and $\alpha=\sqrt{2}$, respectively.

 For both values of the softening parameter, the Soft-Coulomb potential exhibits a strongly non-monotonic dependence of the ionization probability on the field amplitude. After an initial increase at low amplitudes, the ionization probability decreases over a broad interval of field strengths, forming a pronounced stabilization window. This behavior is consistent with the well-known phenomenon of ionization stabilization reported in driven Soft-Coulomb systems.

The existence of a stabilization region indicates that increasing the field strength does not necessarily enhance ionization. Instead, a significant fraction of trajectories remains trapped even under stronger driving fields. For $\alpha=1$, the stabilization window extends over a wide range of amplitudes, producing a substantial reduction in the ionization probability relative to its low-field maximum. A similar behavior is observed for $\alpha=\sqrt{2}$, although the position and depth of the stabilization region are modified by the softening parameter.

A markedly different scenario emerges when the Morse side is introduced. In the Morse-Soft-Coulomb model, the stabilization window is almost completely suppressed for both values of $\alpha$. The ionization probability rapidly approaches unity and remains close to one throughout nearly the entire range of field amplitudes investigated. Consequently, the vast majority of trajectories become ionizing, even in parameter regions where the Soft-Coulomb model exhibits significant stabilization.

The contrasting ionization probabilities observed in the two models suggest that the introduction of the Morse repulsive branch substantially modifies the dynamical mechanisms governing trapping and escape. In particular, the suppression of the stabilization window in the Morse--Soft-Coulomb model indicates that the phase-space structures responsible for long-lived trapping in the Soft-Coulomb system may be strongly reorganized. To investigate this connection and identify the transport structures underlying the different ionization regimes, we next analyze escape-time maps and Lagrangian descriptors over the relevant region of phase space.

\subsection{Escape-time maps and Lagrangian descriptors}
We now analyze the escape dynamics for a field amplitude $F_0=8.0$ and frequency $\omega=0.8$, focusing on the field-free energy shell $E_i=-0.5$ used in the ionization-probability calculations. Two escape-time datasets are shown in the same figure. The background represents the escape time $t_{\rm esc}$ computed over the full two-dimensional $(r_0,p_0)$ grid, whereas the energy shell is colored according to the escape times calculated specifically for initial conditions lying on the shell. Black markers identify nonionizing trajectories from the full phase-space grid, whereas white markers indicate nonionizing trajectories restricted to the energy shell. This representation provides a direct connection between the ionization probabilities and the phase-space dynamics governing trapping and escape.

\begin{figure}[!t]

\begin{subfigure}{0.7\textwidth}
\centering
\includegraphics[
    width=\linewidth
]{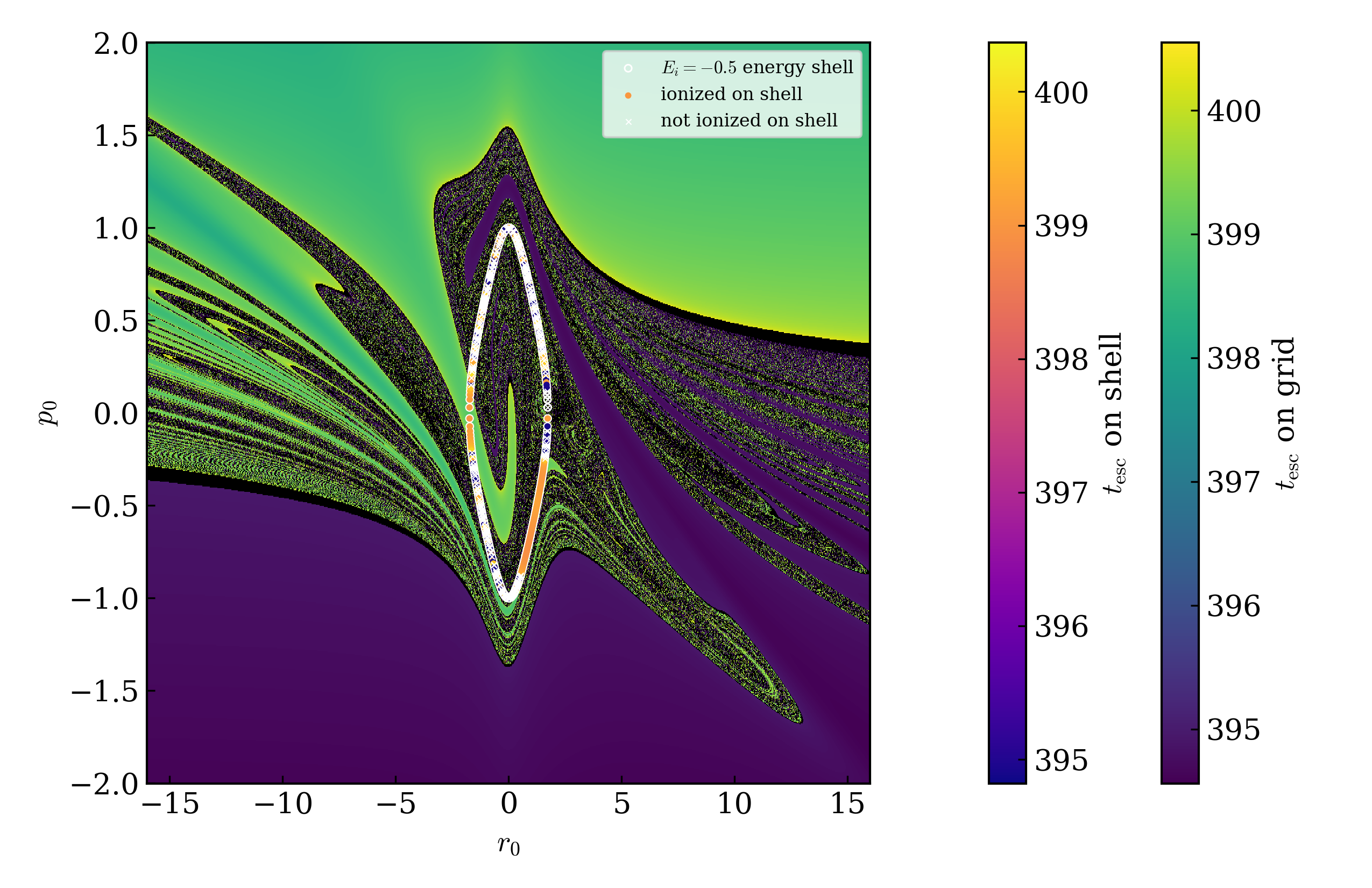}
\caption{Soft-Coulomb}
\end{subfigure}
\hfill
\begin{subfigure}{0.7\textwidth}
\includegraphics[
    width=\linewidth,
]{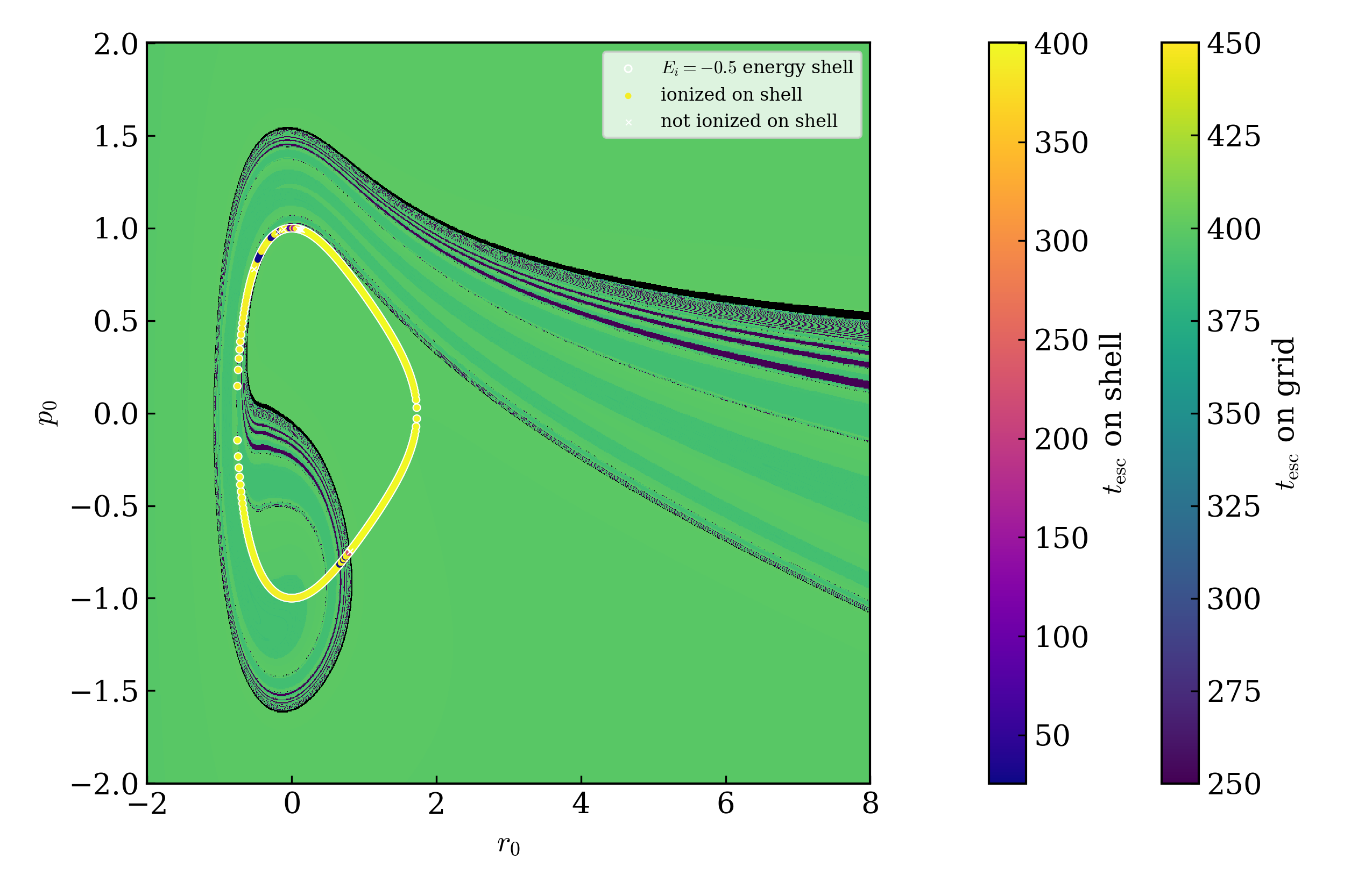}
\caption{Morse--Soft-Coulomb}
\end{subfigure}
\caption{
Escape-time maps for $\alpha=1$, $F_0=8.0$, and $\omega=0.8$:
(a) Soft-Coulomb and (b) Morse--Soft-Coulomb models.
The background color represents $t_{\rm esc}$ computed over the full
$(r_0,p_0)$ grid, whereas the energy shell $E_i=-0.5$ is colored
according to the escape times calculated specifically for initial
conditions on the shell. Black and white markers identify nonionizing trajectories from the full phase-space grid and on the energy shell, respectively.
}
\label{fig:ET_alpha1}
\end{figure}

\begin{figure}[!t]
\centering
\begin{subfigure}{0.7\textwidth}
\includegraphics[width=\linewidth]{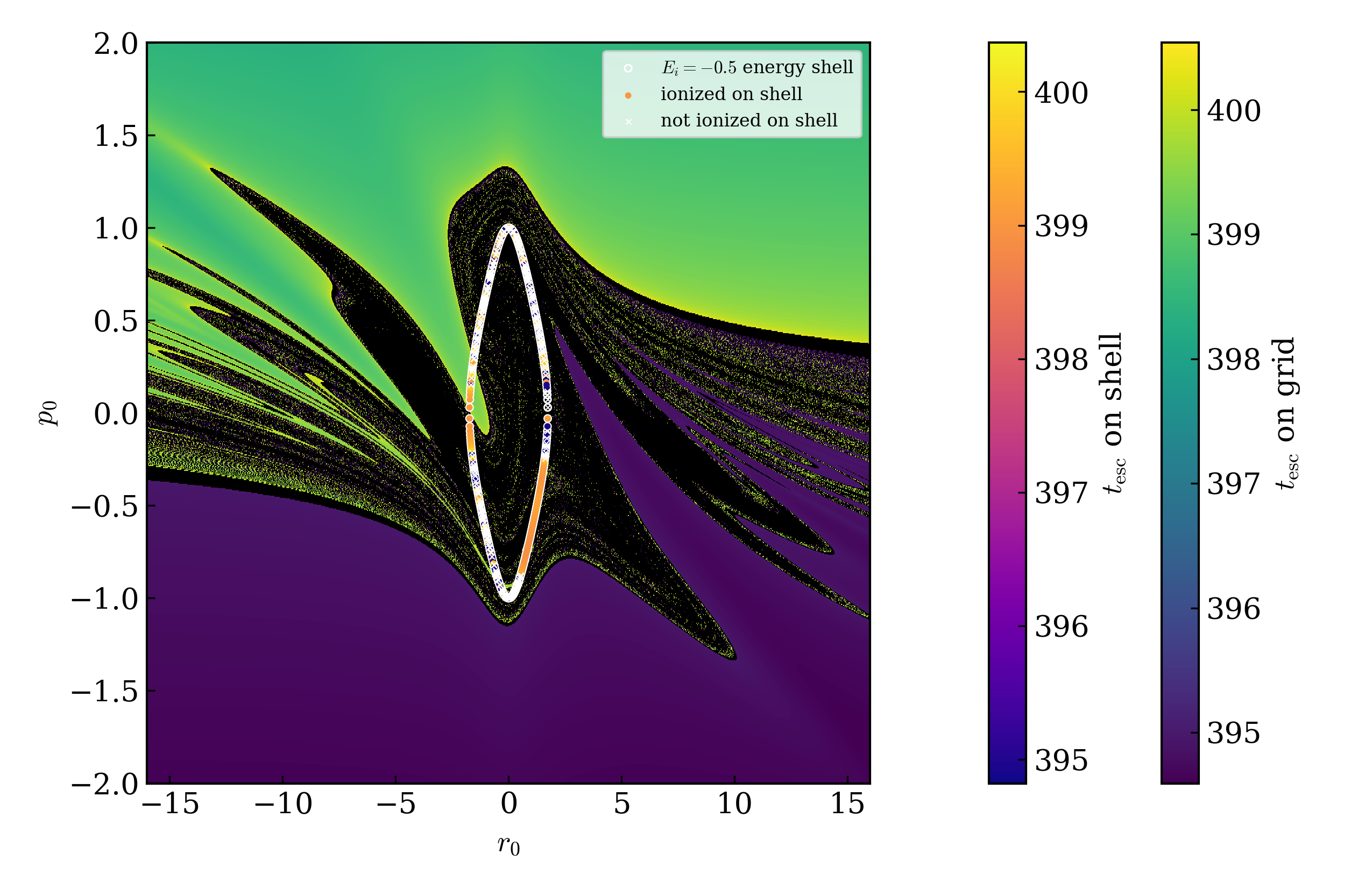}
\caption{Soft-Coulomb}
\end{subfigure}
\hfill
\begin{subfigure}{0.7\textwidth}
\includegraphics[width=\linewidth]{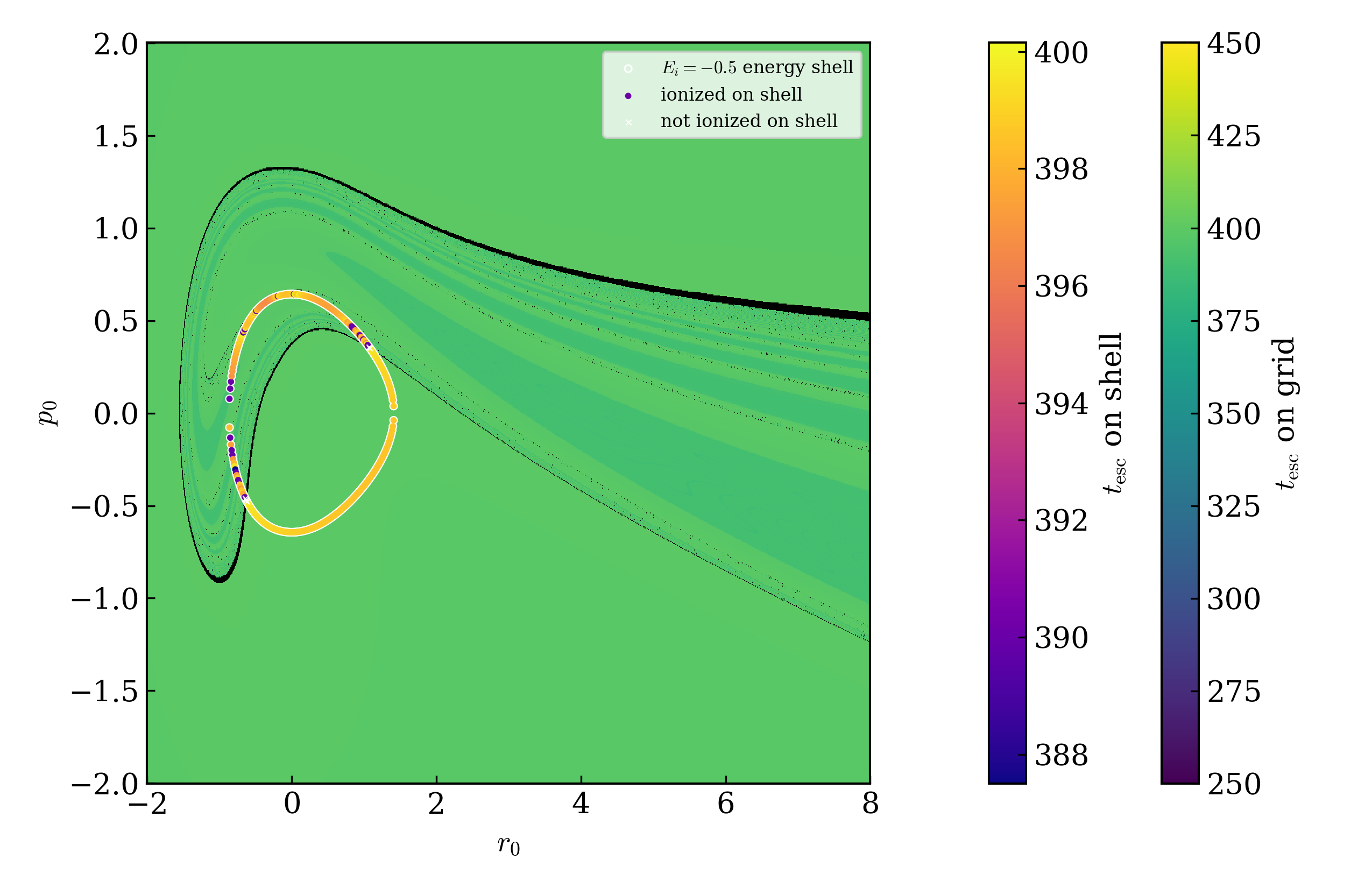}
\caption{Morse--Soft-Coulomb}
\end{subfigure}
\caption{
Escape-time maps for $\alpha=\sqrt{2}$, $F_0=8.0$, and $\omega=0.8$:
(a) Soft-Coulomb and (b) Morse--Soft-Coulomb models.
The background color represents $t_{\rm esc}$ computed over the full
$(r_0,p_0)$ grid, whereas the energy shell $E_i=-0.5$ is colored
according to the escape times calculated specifically for initial
conditions on the shell. Black and white markers identify nonionizing trajectories from the full phase-space grid and on the energy shell, respectively.
}
\label{fig:ET_alpha_sqrt2}
\end{figure}

Figures~\ref{fig:ET_alpha1} and \ref{fig:ET_alpha_sqrt2} reveal a
pronounced difference between the escape dynamics of the Soft-Coulomb
and Morse--Soft-Coulomb models. In the Soft-Coulomb case, the full
phase-space maps display an intricate organization of regions with
different escape times, including narrow filamentary structures and
extended sets of nonionizing initial conditions. The energy shell
$E_i=-0.5$ crosses these dynamically distinct regions, so that
neighboring initial conditions on the same shell may exhibit
substantially different escape times or remain nonionizing throughout
the propagation. This behavior is particularly evident for
$\alpha=1$ [Fig.~\ref{fig:ET_alpha1}(a)], where several nonionizing
segments occur along the shell and the surrounding phase space contains
a complex arrangement of escaping and non-escaping trajectories. For
$\alpha=\sqrt{2}$ [Fig.~\ref{fig:ET_alpha_sqrt2}(a)], the geometry of
these structures and of the energy shell changes, but the coexistence
of ionizing and nonionizing dynamics remains clearly visible.

The Morse--Soft-Coulomb maps exhibit a qualitatively different
organization. For both $\alpha=1$ [Fig.~\ref{fig:ET_alpha1}(b)] and
$\alpha=\sqrt{2}$ [Fig.~\ref{fig:ET_alpha_sqrt2}(b)], the energy shell
is dominated by ionized trajectories, while the
nonionizing portions of the shell are substantially reduced compared
with the corresponding Soft-Coulomb cases. The escape times along the
ionizing portions of the shell are also concentrated within a relatively
narrow interval close to the end of the propagation, despite the
complex structures that remain visible in the surrounding phase space.
Thus, the main effect of replacing the Soft-Coulomb core by the Morse
branch is not the complete disappearance of structured phase-space
dynamics, but a marked change in how the physically relevant energy
shell intersects the regions associated with trapping and escape.

This comparison provides a direct phase-space interpretation of the
ionization-probability results. In the Soft-Coulomb model, the
$E_i=-0.5$ ensemble samples both escaping and non-escaping regions,
allowing a finite fraction of trajectories to remain trapped even in
the strong-field regime. In the Morse--Soft-Coulomb model, the same
energy shell is much less connected to nonionizing regions, and most
of its initial conditions eventually reach the ionizing regime. The
persistence of this distinction indicates that the suppression of the stabilization window is associated with a robust reorganization of the phase-space transport induced by the Morse branch.

To further characterize the phase-space structures underlying the contrasting trapping and escape dynamics revealed by the escape-time maps, we computed Lagrangian descriptors (LDs) over the same phase-space region. For an initial condition $\mathbf{x}_0=(r_0,p_0)$, the LD is defined as the sum of the backward and forward contributions,

\begin{equation}
LD(\mathbf{x}_0)
=
LD^{-}(\mathbf{x}_0)+LD^{+}(\mathbf{x}_0),
\label{eq:LDtotal}
\end{equation}

with

\begin{equation}
LD^{-}(\mathbf{x}_0)
=
\int_{-\tau}^{0}
\left[
|\dot{r}(t;\mathbf{x}_0)|^{\gamma}
+
|\dot{p}(t;\mathbf{x}_0)|^{\gamma}
\right]dt,
\end{equation}

and

\begin{equation}
LD^{+}(\mathbf{x}_0)
=
\int_{0}^{\tau}
\left[
|\dot{r}(t;\mathbf{x}_0)|^{\gamma}
+
|\dot{p}(t;\mathbf{x}_0)|^{\gamma}
\right]dt.
\end{equation}

Here, $\tau$ defines the integration interval in both time directions and $\gamma$ determines the exponent used in the construction of the descriptor. In the present calculations, we set $\tau=100$ and $\gamma=0.4$. The trajectories were propagated using a fourth-order Runge--Kutta scheme with $N_t=40056$ integration steps in each time direction, and the LD integrals were evaluated using the composite Simpson rule. The calculations were performed over the phase-space domain $r_0\in[-15,15]$ and $p_0\in[-2,2]$, with resolutions $\Delta r_0=\Delta p_0=0.01$. To enhance the phase-space structures encoded in the LD, we analyze the magnitude of its gradient,

\begin{equation}
|\nabla LD|
=
\sqrt{
\left(\frac{\partial LD}{\partial r_0}\right)^2
+
\left(\frac{\partial LD}{\partial p_0}\right)^2
},
\label{eq:gradLD}
\end{equation}

where the derivatives are evaluated numerically on the same
phase-space grid. Large spatial variations of the LD produce sharp features in $|\nabla LD|$, providing a sensitive representation of the structures organizing trapping and escape in phase space.

\begin{figure}[!t]
\centering

\begin{subfigure}{0.7\textwidth}
\includegraphics[width=\linewidth]{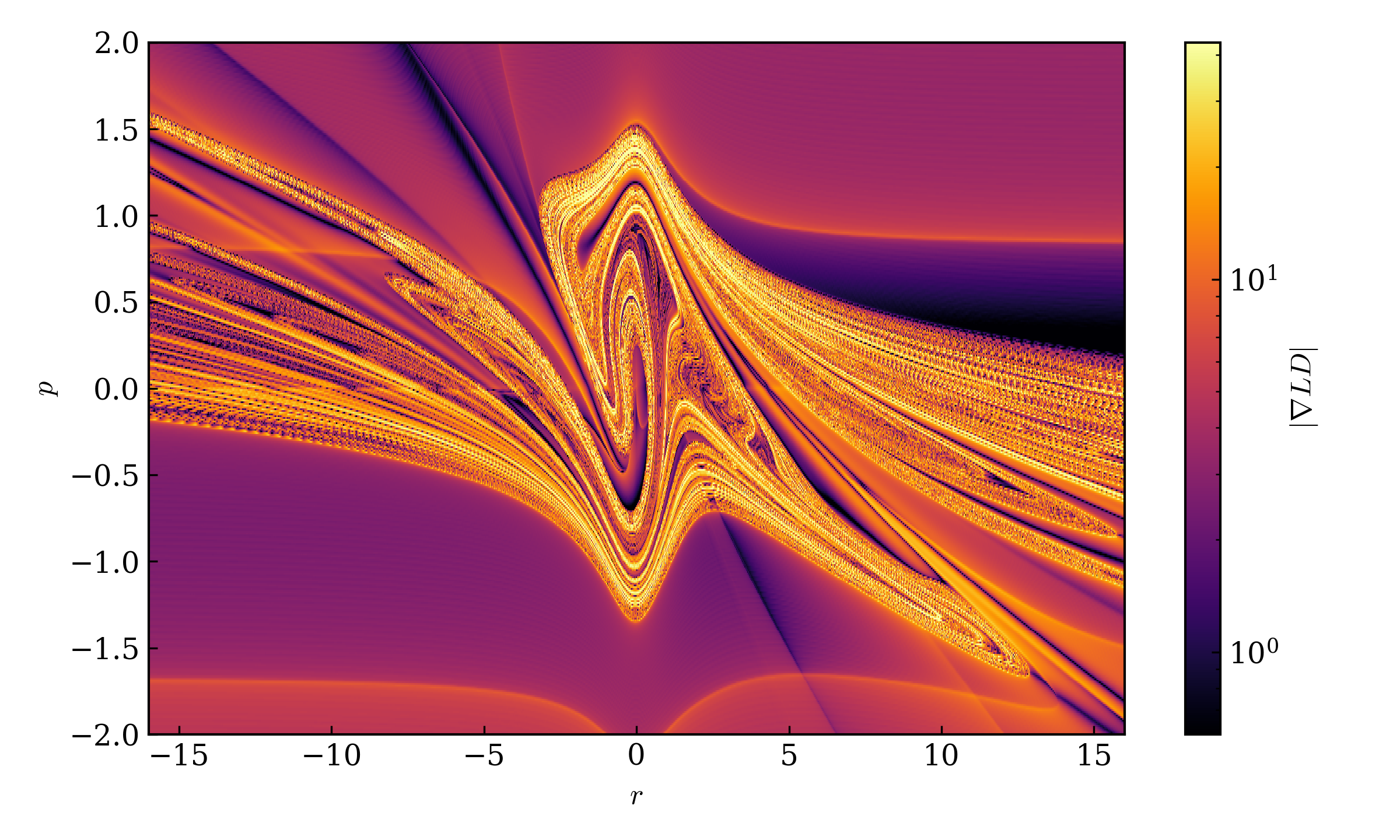}
\caption{Soft-Coulomb}
\end{subfigure}
\hfill
\begin{subfigure}{0.7\textwidth}
\includegraphics[width=\linewidth]{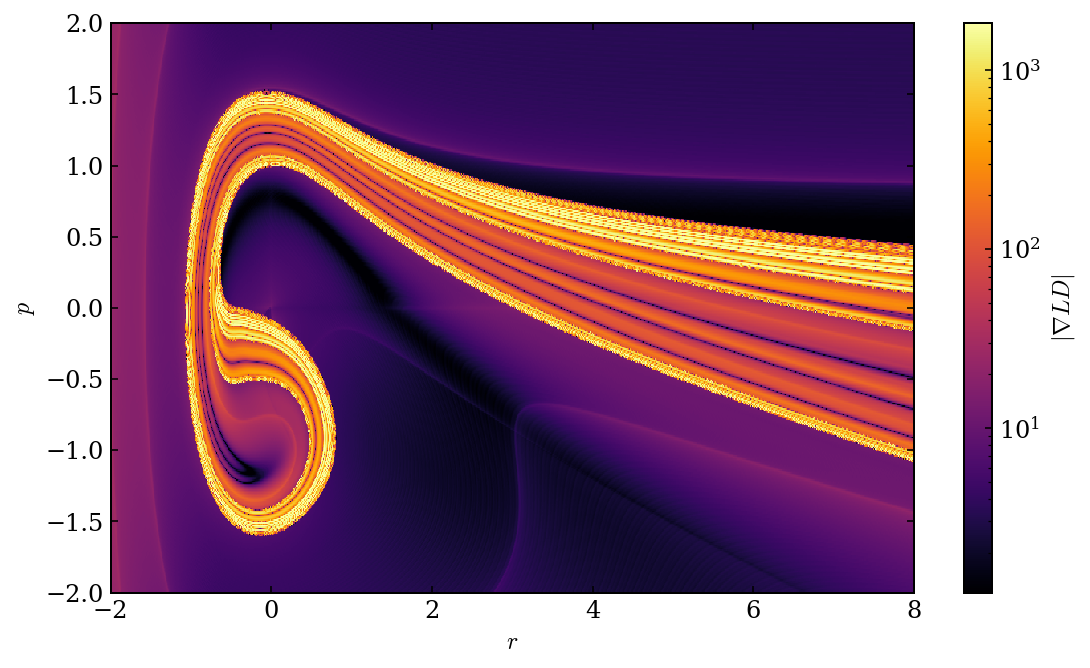}
\caption{Morse--Soft-Coulomb}
\end{subfigure}

\caption{
Gradient magnitude of the Lagrangian descriptor, $|\nabla LD|$,
for (a) the Soft-Coulomb and (b) the Morse--Soft-Coulomb models
with $\alpha=1$, $F_0=8.0$, and $\omega=0.8$.}
\label{fig:LD_alpha1}
\end{figure}

\begin{figure}[!t]
\centering

\begin{subfigure}{0.7\textwidth}
\includegraphics[width=\linewidth]{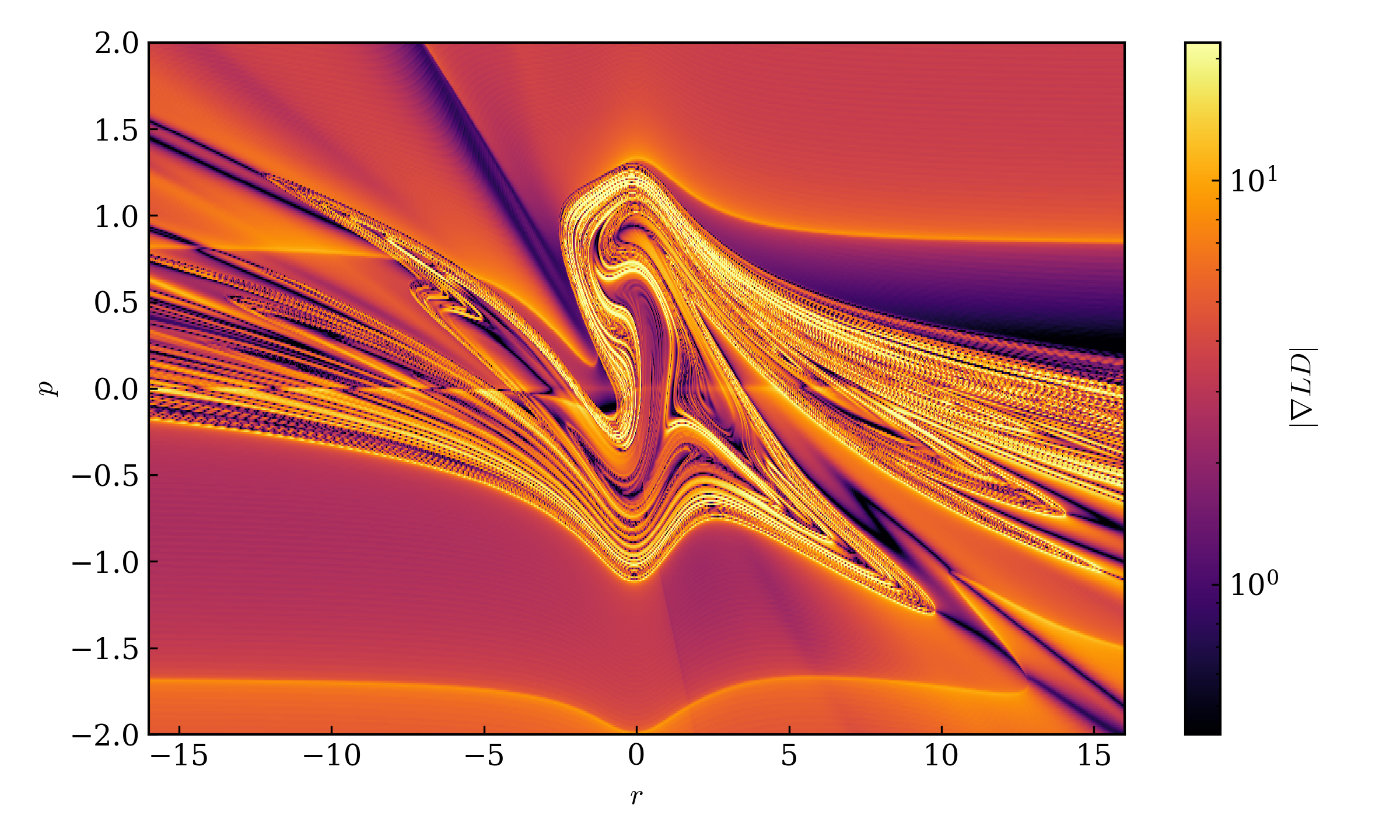}
\caption{Soft-Coulomb}
\end{subfigure}
\hfill
\begin{subfigure}{0.7\textwidth}
\includegraphics[width=\linewidth]{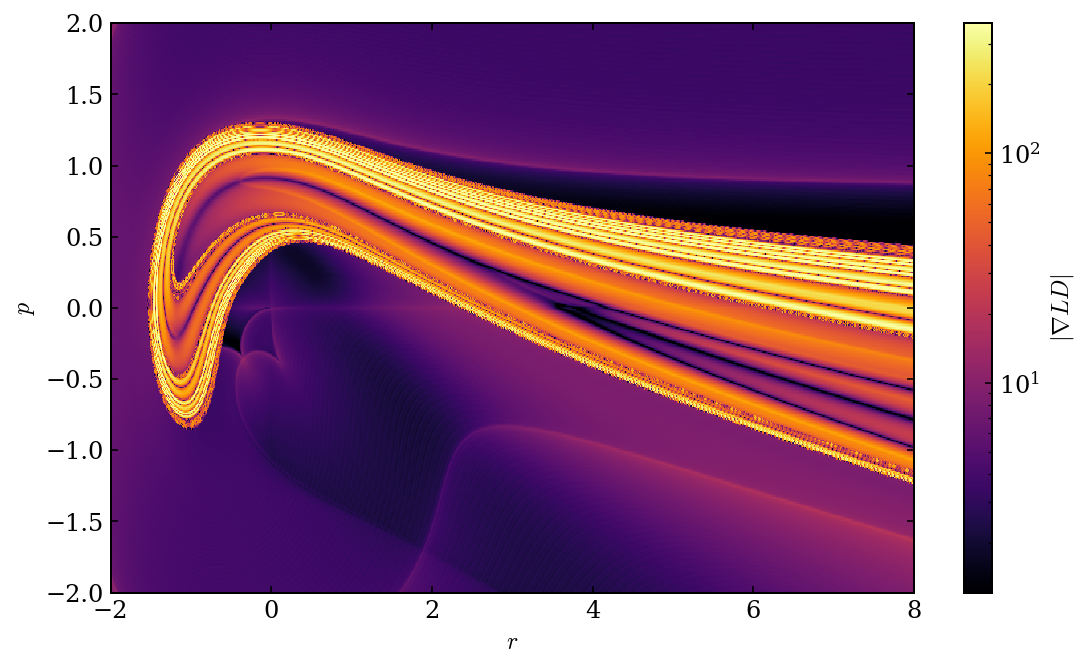}
\caption{Morse--Soft-Coulomb}
\end{subfigure}

\caption{
Gradient magnitude of the Lagrangian descriptor, $|\nabla LD|$,
for (a) the Soft-Coulomb and (b) the Morse--Soft-Coulomb models
with $\alpha=\sqrt2$, $F_0=8.0$, and $\omega=0.8$.}
\label{fig:LD_alpha_sqrt2}
\end{figure}

Figures~\ref{fig:LD_alpha1} and \ref{fig:LD_alpha_sqrt2} show the corresponding LD maps for the Soft-Coulomb and Morse-Soft-Coulomb models. For the Soft-Coulomb potential, the LD reveals a complex network of filamentary structures and sharp ridges distributed throughout phase space. These structures closely follow the regions of long escape times observed in the escape-time maps and delineate transport barriers that separate trapped and escaping trajectories. Large portions of the energy shell intersect these structures, indicating that trajectories initialized on the shell are strongly influenced by the underlying transport skeleton. The presence of these
barriers is consistent with the existence of long trapping times and with the stabilization window observed in the ionization probability curves.

A markedly different scenario is observed in the Morse--Soft-Coulomb model. Although phase-space structures are still present, the LD landscape becomes substantially simpler and less organized than in the Soft-Coulomb case. The intricate network of ridges responsible for delayed transport is strongly reduced, indicating a significant reorganization of the invariant structures governing the dynamics. This simplification is consistent with the predominance of ionizing
trajectories and with the suppression of extended trapping regions observed in the escape-time maps.

The close correspondence between the escape-time maps and the LD structures demonstrates that both diagnostics reveal the same underlying transport skeleton of phase space. Regions characterized by long escape times are systematically associated with LD ridges, confirming that the stabilization observed in the Soft-Coulomb model is closely related to transport barriers that inhibit escape over long time
intervals. The introduction of the Morse branch weakens these barriers, facilitates transport toward ionization, and consequently suppresses the stabilization mechanism.

\subsection{Kramers-Henneberger effective potentials}

To investigate the trapping structures associated with stabilization, we analyze the Kramers-Henneberger (KH) effective potentials corresponding to the Soft-Coulomb and Morse-Soft-Coulomb models.

In the KH frame, the effective potential is obtained by averaging the binding potential over one optical cycle,

\begin{equation}
r \rightarrow r + q\cos(\omega t),
\end{equation}
where $q$ denotes the quiver amplitude of a free electron. The effective KH potential is obtained by averaging the binding potential over one optical cycle,

\begin{equation}
V_{\rm KH}(r)
=
\frac{1}{T}
\int_0^T
V\!\left(r+q\cos(\omega t)\right)\,dt,
\label{eq:vkh}
\end{equation}

Throughout this section, the KH potentials are evaluated for a fixed quiver amplitude $q=F_0/\omega^2=12.5$.

\begin{figure}[!t]
\centering
\includegraphics[width=0.75\textwidth]{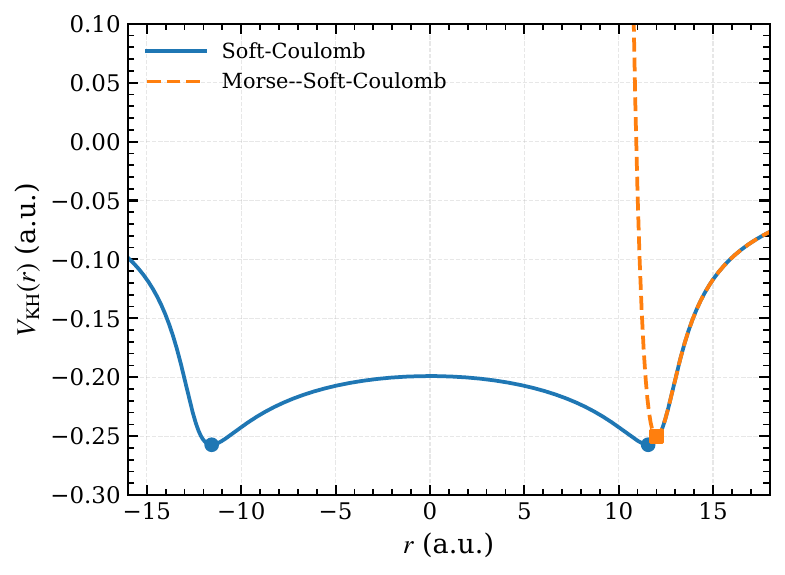}
\caption{
Comparison between the Kramers-Henneberger effective potentials of the Soft-Coulomb and Morse-Soft-Coulomb models for $\alpha=1$ and $q=12.5$. The Morse repulsive branch breaks the symmetry of the Soft-Coulomb double-well structure and suppresses one of the effective trapping minima.
}
\label{fig:KH_comparison}
\end{figure}

Figure~\ref{fig:KH_comparison} compares the KH effective potentials associated with the Soft-Coulomb and Morse-Soft-Coulomb models. The Soft-Coulomb potential develops the well-known symmetric double-well structure, characterized by two equivalent minima located on opposite sides of the origin. Such a structure has long been associated with trapping regions capable of supporting long-lived bounded motion in strongly driven Soft-Coulomb systems.

A markedly different scenario emerges in the Morse-Soft-Coulomb model. The introduction of the Morse repulsive branch breaks the left-right symmetry of the KH potential, suppressing one of the trapping minima and replacing the symmetric double-well structure by a single effective well. As a consequence, the trapping landscape responsible for stabilization is profoundly modified.

The direct comparison between the two KH potentials is presented in Fig.~\ref{fig:KH_comparison}. The symmetry properties of the two effective potentials are fundamentally different. Since the Soft-Coulomb potential is symmetric, its KH counterpart also remains symmetric,$V_{\rm KH}^{\rm SC}(-r)=V_{\rm KH}^{\rm SC}(r)$. As a consequence, the effective potential develops two equivalent trapping minima located symmetrically about the origin. A numerical calculation of the extrema reveals minima at $r_{\rm min}^{\rm SC}\simeq\pm11.564$. In contrast, the Morse-Soft-Coulomb potential is intrinsically asymmetric as well as its KH potential.

The transition from two symmetric minima to a single asymmetric minimum constitutes a direct geometric manifestation of the symmetry breaking induced by the Morse repulsive branch. The disappearance of one of the two trapping wells represents a qualitative modification of the effective landscape experienced by the electron. While the Soft-Coulomb KH potential supports a large trapping region with two minima, the Morse-Soft-Coulomb potential supports only a single minimum, leading to a substantial reduction of the phase-space volume available for trapped motion. The emergence of the double-well structure is not present for all field amplitudes. Numerical calculations performed for different quiver amplitudes reveal that, for sufficiently small values of $q$, the Soft-Coulomb KH potential possesses only a single effective minimum. As the quiver amplitude is increased, a bifurcation occurs and the KH potential develops two symmetric minima separated by a central barrier. The depth and separation of these minima increase with increasing $q$. This behavior is consistent with the ionization-probability curves presented in Figs.~\ref{fig:probability_alpha1} and \ref{fig:probability_alpha_sqrt2}. At low field amplitudes, where the KH potential supports only a single trapping region, the ionization probability remains relatively large. As the field amplitude increases and the double-well structure emerges, additional trapping regions become available in the effective KH dynamics, leading to a reduction of the ionization probability and the onset of the stabilization regime. The appearance of the stabilization window therefore correlates with the formation of the symmetric double-well KH potential and the associated increase in the phase-space volume occupied by bounded trajectories.

It is important to emphasize that the KH potential should be regarded as an effective representation of the trapping landscape rather than a complete dynamical description of stabilization. Recent studies have shown that KH states are more fundamentally associated with periodic orbits and their surrounding phase-space structures in the full
time-dependent Hamiltonian rather than solely with the minima of the cycle-averaged KH potential \cite{Floriani2024KHScars}. In this sense,the KH potential provides an effective visualization of the trapping landscape, while the actual stabilization dynamics is governed by the underlying transport structures of phase space. Nevertheless, the strong modification of the KH potential induced by the Morse repulsive branch provides clear evidence that the trapping region supporting
stabilization is shrunk when the symmetry of the
Soft-Coulomb interaction is broken.

The KH effective potentials therefore provide a dynamical interpretation consistent with the ionization probabilities and escape-time maps. In particular, the Morse repulsive branch strongly breaks the parity symmetry of the effective trapping landscape, eliminating one of the trapping wells and substantially modifying the transport pathways available to the electron.

% ============================================================
\section{Conclusions}
% ============================================================

We have investigated ionization stabilization in driven Soft-Coulomb and Morse--Soft-Coulomb systems through a combined analysis of ionization probabilities, escape-time maps, and Kramers--Henneberger (KH) effective potentials. Our results demonstrate that the introduction of the Morse repulsive branch completely suppresses the stabilization observed in the Soft-Coulomb model, leading to a qualitatively different ionization dynamics.

A central result of the present work is that the suppression of ionization stabilization is governed primarily by the introduction of a repulsive barrier at short range rather than by the Coulomb singularity itself. While previous studies have often associated the disappearance of stabilization with the singular character of the Coulomb potential, our results indicate that the decisive mechanism is the profound reorganization of the phase-space structures responsible for long-lived trapping. From this perspective, the Coulomb singularity can be viewed as a particular realization of a more general repulsive mechanism that inhibits the formation of effective trapping regions under strong laser driving.

This interpretation is consistently supported by all dynamical diagnostics considered in this work. The escape-time and Lagrangian-descriptor maps reveal a substantial reduction of long-lived trapping regions in the Morse--Soft-Coulomb model. The KH analysis further shows that the Soft-Coulomb potential develops a symmetric double-well effective potential, whereas the Morse--Soft-Coulomb potential supports only a single effective minimum.

Interestingly, the behavior observed in the MsC model is qualitatively analogous to that previously reported for higher-dimensional Soft-Coulomb systems, where ionization stabilization becomes significantly weaker or disappears altogether \cite{Menis1992,PhysRevA.46.1638,PhysRevA.46.1638}. We then speculate that this fact may be related to the centrifugal barrier. 

An immediate extension of the present investigation is to determine whether the suppression of stabilization observed at the classical level remains in the quantum MsC counterpart. Such a study may also clarify how the modifications of the KH effective potentials and the associated phase-space structures manifest themselves in the quantum dynamics, providing further insight into the mechanisms responsible for ionization stabilization in intense laser fields.

\bibliographystyle{apsrev4-2}
\bibliography{references}

\end{document}